\documentclass[5p]{elsarticle}

\usepackage[utf8]{inputenc}
\usepackage[T1]{fontenc}
\usepackage{hyperref}
\usepackage{url}
\usepackage{booktabs}
\usepackage{graphicx}
\usepackage{float}
\usepackage{amsfonts}
\usepackage{amsmath}
\usepackage{algorithm}
\usepackage{algorithmic}

\makeatletter
\def\ps@pprintTitle{%
 \let\@oddhead\@empty
 \let\@evenhead\@empty
 \def\@oddfoot{}%
 \let\@evenfoot\@oddfoot}
\makeatother

\begin{document}

\begin{frontmatter}

\title{Matching of Everyday Power Supply and Demand with Dynamic Pricing:\\ Problem Formalisation and Conceptual Analysis}
\author[1]{Thibaut Théate}\corref{cor1}
\ead{thibaut.theate@uliege.be}
\author[2]{Antonio Sutera}
%\ead{antonio.sutera@haulogy.net}
\author[1,3]{Damien Ernst}
\ead{dernst@uliege.be}
\address[1]{Department of Electrical Engineering and Computer Science, University of Liège, Liège, Belgium}
\address[2]{Haulogy, Intelligent Systems Solutions, Braine-le-Comte, Belgium}
\address[3]{Information Processing and Communications Laboratory, Institut Polytechnique de Paris, Paris, France}
\cortext[cor1]{Corresponding author.}

\begin{abstract}

    The energy transition is expected to significantly increase the share of renewable energy sources whose production is intermittent in the electricity mix. Apart from key benefits, this development has the major drawback of generating a mismatch between power supply and demand. The innovative \textit{dynamic pricing} approach may significantly contribute to mitigating that critical problem by taking advantage of the flexibility offered by the demand side. At its core, this idea consists in providing the consumer with a price signal which is evolving over time, in order to influence its consumption. This novel approach involves a challenging decision-making problem that can be summarised as follows: how to determine a price signal maximising the synchronisation between power supply and demand under the constraints of maintaining the producer/retailer's profitability and benefiting the final consumer at the same time? As a contribution, this research work presents a detailed formalisation of this particular decision-making problem. Moreover, the paper discusses the diverse algorithmic components required to efficiently design a dynamic pricing policy: different forecasting models together with an accurate statistical modelling of the demand response to dynamic prices.
    
\end{abstract}


\end{frontmatter}

\section{Introduction}
\label{SectionIntroduction}

Climate change is undeniably a major challenge facing humanity in the 21st century \cite{IPCC2021}. An ambitious transformation is required in all sectors to significantly lower their respective carbon footprints. Electricity generation is no exception, with the burning of fossil fuels, mainly coal and gas, being by far the dominant power source in the world today \cite{Ritchie2020}. This sector has to undergo an important transformation of the global electricity mix by promoting power sources with a significantly lower carbon footprint. Belonging to that category are nuclear power, hydroelectricity, biomass or geothermal energy which are relatively controllable, but also the energy directly extracted from wind and sun which is conversely intermittent in nature. Since wind turbines and photovoltaic panels are expected to play a key role in the energy transition, solutions are required to address their variable production. Interesting technical avenues are the interconnection of power grids \cite{Chatzivasileiadis2012} and the development of storage capacities such as battery, pumped hydroelectricity or hydrogen \cite{Kittner2017}. Another promising and innovative solution is to influence the behaviour of consumers through the use of \textit{dynamic pricing} (DP), so that power supply and demand are better synchronised.\\

\begin{figure}
    \centering
    \includegraphics[width=1\linewidth, trim={7.5cm 3.3cm 9cm 3.2cm}, clip]{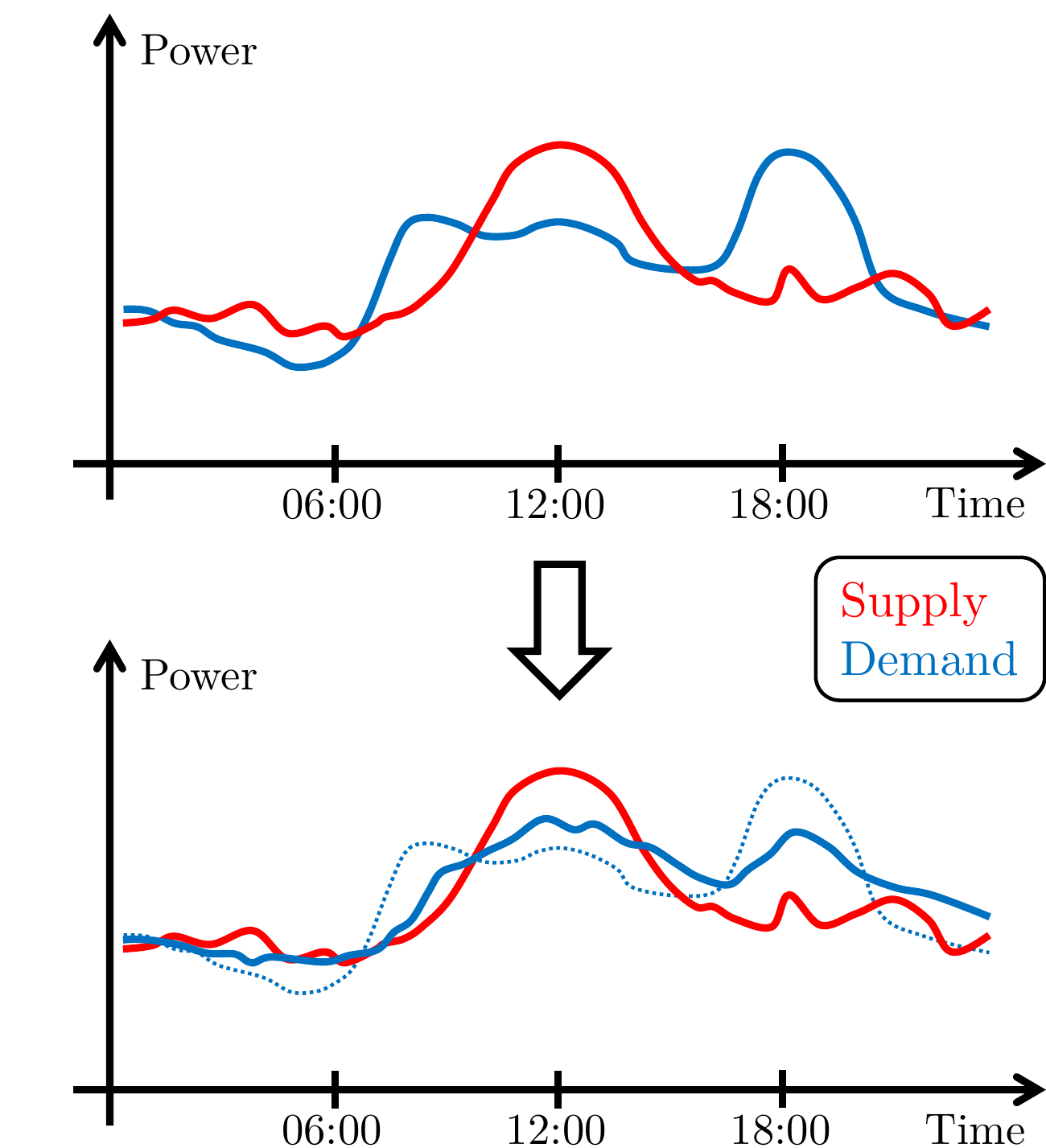}
    \caption{Illustration of the dynamic pricing approach's potential to shift and change the shape of a typical daily consumption curve (blue) so that there is a better synchronisation with the daily intermittent production curve of renewable energy sources (red).}
    \label{FigureDemandResponseDynamicPricing}
\end{figure}

The dynamic pricing approach consists in continuously adapting the electricity price that the final consumer has to pay in order to influence its consumption behaviour. Basically, when demand exceeds supply, the power price would be increased in order to take down consumption. Conversely, a reduced price would be provided when there is excessive production compared to consumption. From a graphical perspective, the objective is not only to shift the daily consumption curve but also to change its shape in order to better overlap with the intermittent production curve of renewable energy sources. This is illustrated in Figure \ref{FigureDemandResponseDynamicPricing} for a representative situation.\\

The innovative dynamic pricing approach relies on two important assumptions. Firstly, the final consumer has to be equipped with a smart metering device to measure its production in real-time and with communication means for the price signal. Secondly, the final consumer has to be able to provide a certain amount of flexibility regarding its power consumption. Moreover, it has to be sufficiently receptive to the incentives offered to reduce its electricity bill in exchange for a behaviour change. If these requirements are met, the major strength of the dynamic pricing approach is its potential benefits for both the consumer and the producer/retailer. Moreover, these benefits would not only be in terms of economy, but also potentially in terms of ecology and autonomy. In fact, dynamic prices reward the flexibility of the demand side.\\

The contributions of this research work are twofold. Firstly, the complex decision-making problem faced by a producer/retailer willing to develop a dynamic pricing strategy is presented and rigorously formalised. Secondly, the diverse algorithmic components required to efficiently design a dynamic pricing policy are thoroughly discussed. To the authors' knowledge, demand response via dynamic pricing has received considerable attention from the research community, but from the perspective of the demand side alone. Therefore, the present research may be considered as a pioneer work studying dynamic pricing from the perspective of the supply side for taking advantage of the flexibility of the power consumers.\\

This research paper is structured as follows. First of all, the scientific literature about both dynamic pricing and demand response is concisely reviewed in Section \ref{SectionLiterature}. Then, Section \ref{SectionFormalisation} presents a detailed formalisation of the decision-making problem behind the novel dynamic pricing approach. Afterwards, Section \ref{SectionDiscussion} analyses the algorithmic components necessary for the development of dynamic pricing policies. Subsequently, a fair performance assessment methodology is introduced in Section \ref{SectionPerformanceAssessment} to quantitatively evaluate the performance of a dynamic pricing policy. To end this paper, Section \ref{SectionConclusion} discusses interesting research avenues for future work and draws conclusions.

\section{Literature review}
\label{SectionLiterature}

Over the last decade, the management of the demand side in the scope of the energy transition has received increasing attention from the research community. In fact, there exist multiple generic approaches when it comes to demand response. Without getting into too many details, the scientific literature includes some surveys summarising and discussing the different techniques available together with their associated challenges and benefits \cite{Palensky2011, Siano2014, Deng2015, Vardakas2015, Haider2016}. In this research work, the focus is exclusively set on the demand response induced by dynamic power prices.\\

As previously mentioned, the scientific literature about demand response via dynamic pricing is primarily focused on the perspective of the demand side. Multiple techniques have already been proposed to help the consumer provide flexibility and take advantage of behavioural changes to lower its electricity bill. For instance, \cite{Zhao2013} presents a power scheduling method based on a genetic algorithm to optimise residential demand response via an energy management system, so that the electricity cost is reduced. In \cite{Muratori2016}, a technique based on dynamic programming is introduced for determining the optimal schedule of residential controllable appliances in the context of time-varying power pricing. One can also mention \cite{Liu2017} that proposes an energy sharing model with price-based demand response for microgrids of peer-to-peer prosumers. The approach is based on a distributed iterative algorithm and has been shown to lower the prosumers' costs and improve the sharing of photovoltaic energy. More recently, (deep) reinforcement learning techniques have been proven to be particularly relevant for controlling the residential demand response in the context of dynamic power prices \cite{VazquezCanteli2019, Li2020}.\\

On the contrary, the question of inducing a residential demand response based on a dynamic pricing approach from the perspective of the supply side has not received a lot of attention from the research community yet. Still, there are a few works in the scientific literature about the mathematical modelling of the demand response caused by dynamic power prices, which is a key element in achieving that objective. To begin with, \cite{Gottwalt2011} presents a simulation model highlighting the evolution of electricity consumption profiles when shifting from a fixed tariff to dynamic power prices. The same objective is pursued by \cite{Bregere2020} which introduces a fully data-driven approach relying on the data collected by smart meters and exogenous variables. The resulting simulation model is based on consumption profiles clustering and conditional variational autoencoders. Alternatively, \cite{Ganesan2022} presents a functional model of residential power consumption elasticity under dynamic pricing to assess the impact of different electricity price levels, based on a Bayesian probabilistic approach. In addition to these mathematical models, one can also mention some real-life experiments conducted to assess the responsiveness of residential electricity demand to dynamic pricing \cite{He2012, Klaassen2016}.

\section{Problem formalisation}
\label{SectionFormalisation}

This section presents a mathematical formalisation of the challenging sequential decision-making problem related to the dynamic pricing approach for inducing a residential demand response. To begin with, the contextualisation considered for studying this particular problem is briefly described, followed by an overview of the decision-making process. Then, a discretisation of the continuous timeline is introduced. Subsequently, the formal definition of a dynamic pricing policy is presented. Lastly, the input and output spaces of a dynamic pricing policy are described, together with the objective criterion.

\subsection{Contextualisation}
\label{SectionContextualisation}

As previously hinted, this research work focuses on the interesting real-case scenario of a producer/retailer whose production portfolio is composed of an important share of renewable energy sources such as wind turbines and photovoltaic panels. Because of the substantial intermittency of these generation assets, a strong connection to the energy markets is required in order to fully satisfy its customers regardless of the weather. Nevertheless, the consumers are assumed to be well informed and willing to adapt their behaviour in order to consume renewable energy rather than electricity purchased on the market whose origin may be unknown. Within this particular context, the benefits of the dynamic pricing approach taking advantage of the consumers' flexibility are maximised. Indeed, the insignificant marginal cost associated with these intermittent renewable energy sources coupled with their low carbon footprint make this innovative approach interesting from an economical perspective for both supply and demand sides, but also in terms of ecology. Moreover, the autonomy of the producer/retailer is expected to be reinforced by lowering its dependence on the energy markets. At the same time, dependence on fossil fuels may be reduced as well.\\

In this research work, the predicted difference between power production and consumption is assumed to be fully secured in the day-ahead electricity market. Also called spot market, the day-ahead market has an hourly resolution and is operated once a day for all hours of the following day via a single-blind auction. In other words, trading power for hour $H$ of day $D$ has to be performed ahead on day $D-1$ between 00:00 AM (market opening) and 12:00 AM (market closure). Therefore, the energy is at best purchased 12 hours (00:00 AM of day $D$) up to 35 hours (11:00 PM of day $D$) before the actual delivery of power. Apart from the day-ahead electricity market, it is assumed that there are no trading activities on the future/forward nor intraday markets. Nevertheless, if there remains an eventual mismatch between production and consumption at the time of power delivery, the producer/retailer would be exposed to the imbalance market. In this case, the so-called imbalance price has to be inevitably paid as compensation for pushing the power grid off balance.

\subsection{Decision-making process overview}
\label{SectionDecisionMaking}

The decision-making problem studied in this research work is characterised by a particularity: a variable time lag between the moment a decision is made and the moment it becomes effective. As previously explained, any remaining difference between production and consumption after demand response has to ideally be traded on the day-ahead market. The purpose of this assumption is to limit the exposure of the producer/retailer to the imbalance market. For this reason, the price signal sent to the consumer on day $D$ has to be generated before the closing of the day-ahead market on day $D-1$. Additionally, it is assumed that the price signal cannot be refreshed afterwards.\\

Basically, the decision-making problem at hand can be formalised as follows. The core objective is to determine a decision-making policy, denoted $\Pi$, mapping at time $\tau$ input information of diverse nature $I_\tau$ to the electricity price signal $S_\tau$ to be sent to the consumers over a future time horizon well-defined:
\begin{equation}
    S_\tau = \Pi(I_\tau) \text{,}
\end{equation}
where:
\begin{itemize}
    \item [$\bullet$] $I_\tau$ represents the information vector gathering all the available information (of diverse nature) at time $\tau$ which may be helpful to make a relevant dynamic pricing decision,
    \item [$\bullet$] $S_\tau$ represents a set of electricity prices generated at time $\tau$ and shaping the dynamic price signal over a well-defined future time horizon.
\end{itemize}

The dynamic pricing approach from the perspective of the supply side belongs to a particular class of decision-making problems: automated planning and scheduling. Contrarily to conventional decision-making outputting one action at a time, planning decision-making is concerned with the generation of a sequence of actions. In other words, a planning decision-making problem requires to synthesise in advance a strategy or plan of actions over a certain time horizon. Formally, the decision-making has to be performed at a specific time $\tau$ about a control variable over a future time horizon beginning at time $\tau_i > \tau$ and ending at time $\tau_f > \tau_i$. In this case, the decision-making is assumed to be performed just before the closing of the day-ahead market at 12:00 AM to determine the price signal to be sent to the consumers throughout the entire following day (from 00:00 AM to 11:59 PM).\\

In the next sections, a more accurate and thorough mathematical formalisation of the dynamic pricing problem from the perspective of the supply side is presented. Moreover, the planning problem previously introduced is cast into a sequential decision-making problem. Indeed, this research paper intends to focus on a decision-making policy outputting a single price from the signal $S_\tau$ at a time based on a subset of the information vector $I_\tau$.

\begin{figure*}
    \centering
    \includegraphics[width=0.875 \linewidth, trim={4cm 3.8cm 4cm 4.3cm}, clip]{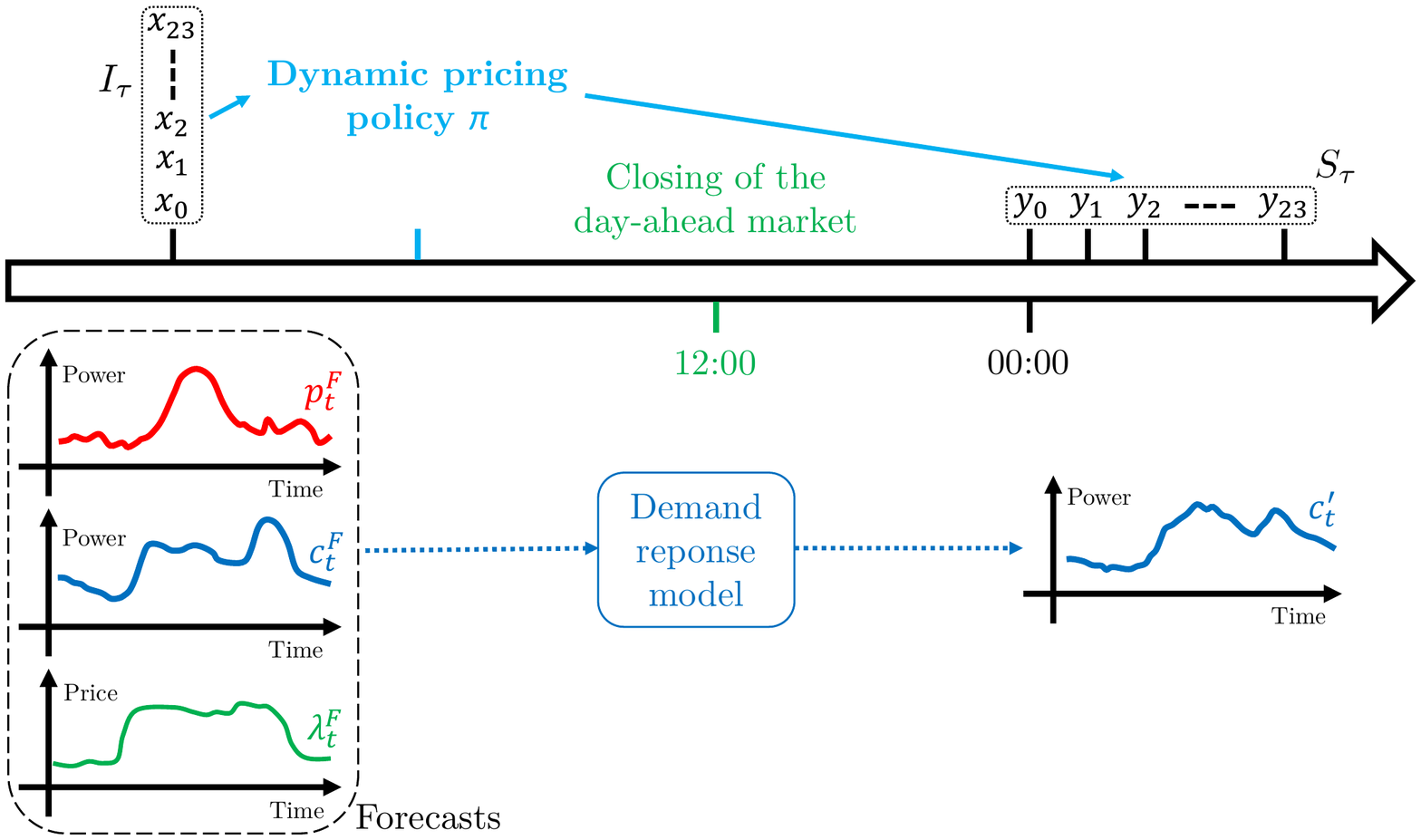}
    \caption{Illustration of the formalised decision-making problem related to dynamic pricing from the perspective of the supply side. The notations $x_t$ and $y_t$ represent the inputs and outputs of a dynamic pricing policy $\pi$, which are not shown concurrent on the timeline since the decision-making occurs multiple hours before the application of the dynamic pricing signal. The time axis of the four plots represents the complete following day for which the dynamic prices are generated. The mathematical notations $p_t^F$, $c_t^F$ and $\lambda_t^F$ respectively represent the forecast production, consumption and day-ahead market price for the time step $t$. Finally, the quantity $c_t'$ is the predicted consumption at time step $t$ after taking into consideration the dynamic pricing signal.}
    \label{FigureProblemFormalisation}
\end{figure*}

\subsection{Timeline discretisation}
\label{SectionDiscretisation}

Theoretically, the dynamic electricity price signal sent to the consumer could be continuously changing over time. More realistically, this research work adopts a discretisation of the continuous timeline so that this power price is adapted at regular intervals. Formally, this timeline is discretised into a number of time steps $t$ spaced by a constant duration $\Delta t$. If the duration $\Delta t$ is too large, the synchronisation improvement between supply and demand will probably be of poor quality. Conversely, lowering the value of the duration $\Delta t$ increases the complexity of the decision-making process, and a too high update frequency may even confuse the consumer. There is a trade-off to be found concerning this important parameter. In this research work, the dynamic price signal is assumed to change once per hour, meaning that $\Delta t$ is equal to one hour. This choice is motivated by the hourly resolution of the day-ahead market, which has proven to be an appropriate compromise over the years for matching power production and consumption. Another relevant discretisation choice could be to have a price signal which is updated every quarter of an hour. In the rest of this research paper, the increment (decrement) operations $t+1$ ($t-1$) are used to model the discrete transition from time step t to time step $t + \Delta t$ ($t - \Delta t$), for the sake of clarity.

\subsection{Dynamic pricing policy}
\label{SectionPolicy}

Within the context previously described, a dynamic pricing planning policy $\Pi$ consists of the set of rules used to make a decision regarding the future price signal sent to the consumers over the next day. This planning policy can be decomposed into a set of 24 dynamic pricing decision-making policies $\pi$ outputting a single electricity price for one hour of the following day. Mathematically, such a dynamic pricing strategy can be defined as a programmed policy $\pi : \mathcal{X} \rightarrow \mathcal{Y}$, either deterministic or stochastic, which outputs a decision $y_t \in Y$ for time step $t$ based on some input information $x_t \in \mathcal{X}$ so as to maximise an objective criterion. The input $x_t$ is derived from the information vector $I_\tau$ associated with the decision-making for time step $t$, after potential preprocessing operations. The price signal $S_\tau$ is composed of 24 dynamic pricing policy outputs $y_t$.\\

In the rest of this research work, the time at which the decision-making does occur should not be confused with the time at which the dynamic price signal is active (charging for energy consumption). The proposed formalisation assumes that the time step $t$ refers to the time at which the dynamic price is active, not decided. Therefore, the decision-making of the dynamic pricing policy for time step $t$ ($y_t = \pi(x_t)$) is in fact performed hours in advance of time step $t$. This complexity is illustrated in Figure \ref{FigureProblemFormalisation} describing the formalised decision-making problem.

\subsection{Input of a dynamic pricing policy}
\label{SectionInput}

The input space $\mathcal{X}$ of a dynamic pricing policy $\pi$ comprises all the available information which may help to make a relevant decision about future electricity prices so that an appropriate demand response is induced. Since the decision-making occurs 12 up to 35 hours in advance of the price signal delivery, this information mainly consists of forecasts and estimations that are subject to uncertainty. As depicted in Figure \ref{FigureProblemFormalisation}, the dynamic pricing policy input $x_t \in \mathcal{X}$ refers to the decision-making occurring at time $\tau = t-h$ with $h \in [12,\ 35]$ about the dynamic pricing signal delivered to the consumer at time step $t$. In fact, the quantity $I_\tau$ may be seen as the information contained in the 24 inputs $x_t$ for $t \in \{\tau+12, ..., \tau+35 \}$. Formally, the input $x_t \in \mathcal{X}$ is decided to be defined as follows:
\begin{equation}
    x_t = \{P_t^{F},\ C_t^{F},\ \Lambda_t^{F},\ Y_t,\ \mathcal{M} \} \text{,}
\end{equation}
where:
\begin{itemize}
    \item [$\bullet$] $P_t^{F} = \{p_{t + \epsilon}^{F} \in \mathbb{R}^+ \ | \ \epsilon = -k, ..., k\}$ represents a set of forecasts for the power production within a time window centred around time step $t$ and of size $k$,
    \item [$\bullet$] $C_t^{F} = \{c_{t + \epsilon}^{F} \in \mathbb{R}^+ \ | \ \epsilon = -k, ..., k\}$ represents a set of forecasts for the power consumption within a time window centred around time step $t$ and of size $k$,
    \item [$\bullet$] $\Lambda_t^{F} = \{\lambda_{t + \epsilon}^{F} \in \mathbb{R} \ | \ \epsilon = -k, ..., k\}$ represents a set of forecasts for the day-ahead market prices within a window centred around time step $t$ and of size $k$,
    \item [$\bullet$] $Y_t = \{y_{t - \epsilon} \in \mathbb{R} \ | \ \epsilon = 1, ..., k\}$ represents the series of $k$ previous values for the dynamic price signal sent to the final consumer,
    \item [$\bullet$] $\mathcal{M}$ is a mathematical model of the demand response to be expected from the consumption portfolio, with the required input information.
\end{itemize}

The different forecasting models and the challenging modelling of the consumption portfolio demand response are discussed in more details in Section \ref{SectionDiscussion}.

\subsection{Output of a dynamic pricing policy}
\label{SectionOutput}

The output space $\mathcal{Y}$ of a dynamic pricing policy $\pi$ only includes the future price signal to be sent to the consumer. Formally, the dynamic pricing policy output $y_t \in \mathcal{Y}$, which represents the electricity price to be paid by the consumer for its power consumption at time step $t$, is mathematically defined as follows:
\begin{equation}
    y_t = e_t \text{,}
\end{equation}
where $e_t \in \mathbb{R}$ represents the dynamic electricity price to be paid by the demand side for its power consumption at time step $t$. Out of the scope of this research work is the presentation of this price signal so that the impact on the final consumer is maximised. Indeed, the way of communicating the output of the dynamic pricing policy has to be adapted to the audience, be it humans with different levels of electricity market expertise or algorithms (energy management systems).

\subsection{Objective criterion}
\label{SectionObjective}

The dynamic pricing approach can provide multiple benefits, in terms of economy, ecology but also autonomy. Consequently, the objective criterion to be maximised by a dynamic pricing policy $\pi$ is not trivially determined. In fact, several core objectives can be clearly identified:
\begin{itemize}
    \item [$\bullet$] maximising the match between supply and demand,
    \item [$\bullet$] minimising the carbon footprint of power generation,
    \item [$\bullet$] minimising the electricity costs for the consumer,
    \item [$\bullet$] maximising the revenue of the producer/retailer.
\end{itemize}

Although some objectives overlap, these four criteria are not completely compatible. For instance, maximising the synchronisation between power supply and demand is equivalent to minimising the carbon footprint associated with the generation of electricity. Indeed, the production portfolio of the producer/retailer being mainly composed of intermittent renewable energy sources, its energy has a reduced carbon footprint compared to the electricity that can be purchased on the day-ahead market whose origin is unknown. On the contrary, maximising the revenue of the producer/retailer will obviously not lead to a minimised electricity bill for the consumer. This research work makes the choice to prioritise the maximisation of the synchronisation between supply and demand, and equivalently the minimisation of the carbon footprint, while translating the other two core objectives into relevant constraints. Firstly, the costs for the consumer have to be reduced with respect to the situation without dynamic pricing. Secondly, the profitability of the producer/retailer has to be guaranteed.\\

Formally, the objective criterion to be optimised by a dynamic pricing policy $\pi$ can be mathematically defined as the following. First of all, the main target to evaluate is the synchronisation between supply and demand, which can be quantitatively assessed through the deviation $\Delta_T$. This quantity has to ideally be minimised, and can be mathematically expressed as follows:
\begin{equation}
    \Delta_T = \sum_{t=0}^{T-1} |p_t - c_t| \text{,}
\end{equation}
where:
\begin{itemize}
    \item [$\bullet$] $t=0$ corresponds to the first electricity delivery hour of a new day (00:00 AM),
    \item [$\bullet$] $T$ is the time horizon considered, which should be a multiple of 24 to have full days,
    \item [$\bullet$] $p_t$ is the actual power production (not predicted) from the supply side at time step $t$,
    \item [$\bullet$] $c_t$ is the actual power consumption (not predicted) from the demand side at time step $t$.
\end{itemize}

Afterwards, the first constraint concerning the reduced costs for the consumer has to be modelled mathematically. This is achieved via the electricity bill $B_T$ paid by the consumer over the time horizon $T$, which can be expressed as the following:
\begin{equation}
    B_T = \sum_{t=0}^{T-1} c_t\ y_t \ \text{.}
\end{equation}

As previously explained, the consumer power bill $B_T$ should not exceed that obtained without dynamic pricing. In that case, the consumer is assumed to pay a price $\overline{e_t}$, which can for instance be a fixed tariff or a price indexed on the day-ahead market price. The situation without dynamic pricing is discussed in more details in Section \ref{SectionPerformanceAssessment}. Consequently, the first constraint can be mathematically expressed as follows:
\begin{equation}
    \sum_{t=0}^{T-1} c_t\ y_t \le \sum_{t=0}^{T-1} \overline{c_t}\ \overline{e_t} \ \text{,}
\end{equation}
where $\overline{c_t}$ is the power consumption from the demand side at time step $t$ without dynamic pricing.\\

Then, the second constraint is about the profitability of the producer/retailer, which is achieved if its revenue exceeds its costs. The revenue $R_T$ of the producer/retailer over the time horizon $T$ can be mathematically expressed as the following:
\begin{equation}
    R_T = \sum_{t=0}^{T-1} \left[ c_t\ y_t - (c_t' - p_t^F) \ \lambda_t - (c_t - p_t) \ i_t \right] \text{,}
\end{equation}
where:
\begin{itemize}
    \item [$\bullet$] $\lambda_t$ is the actual power price (not predicted) on the day-ahead market at time step $t$,
    \item [$\bullet$] $i_t$ is the actual imbalance price (not predicted) on the imbalance market at time step $t$,
    \item [$\bullet$] $c_t'$ is the predicted power consumption at time step $t$ after demand response to the dynamic prices, based on the demand response mathematical model $\mathcal{M}$.
\end{itemize}

The first term corresponds to the payment of the customers for their electricity consumption. The second term is the revenue or cost induced by the predicted mismatch between supply and demand, which is traded on the day-ahead market. The last term is the cost or revenue caused by the remaining imbalance between supply and demand, which has to be compensated in the imbalance market.\\

The total costs incurred by the producer/retailer at each time step $t$ can be decomposed into both fixed costs $F_C$ and marginal costs $M_C$. In this particular case, the marginal costs of production are assumed to be negligible since the production portfolio is composed of intermittent renewable energy sources such as wind turbines and photovoltaic panels. Therefore, the second constraint can be mathematically expressed as follows:
\begin{equation}
    \sum_{t=0}^{T-1} \left[ c_t\ y_t - (c_t' - p_t^F) \ \lambda_t - (c_t - p_t) \ i_t \right] \ge F_C\ T \ \text{.}
\end{equation}

Finally, the complete objective criterion to be optimised by a dynamic pricing policy can be mathematically expressed as follows:
\begin{equation}
\begin{aligned}
& \underset{\pi}{\text{minimise}}
& & \sum_{t=0}^{T-1} |p_t - c_t| \text{,} \\
& \text{subject to}
& & R_T \ge F_C\ T \ \text{,} \\
& & & B_T \le \sum_{t=0}^{T-1} \overline{c_t}\ \overline{e_t} \ \text{.}
\end{aligned}
\end{equation}

\section{Algorithmic components discussion}
\label{SectionDiscussion}

This section presents a thorough discussion about the different algorithmic modules required to efficiently design a dynamic pricing policy from the perspective of the supply side. Firstly, the different forecasting blocks are rigorously analysed. Secondly, the modelling of the demand response induced by dynamic prices is discussed. Lastly, the proper management of uncertainty is considered.\\

\begin{figure*}
    \centering
    \includegraphics[width=0.9\linewidth, trim={4.7cm 2.7cm 3.7cm 2.7cm}, clip]{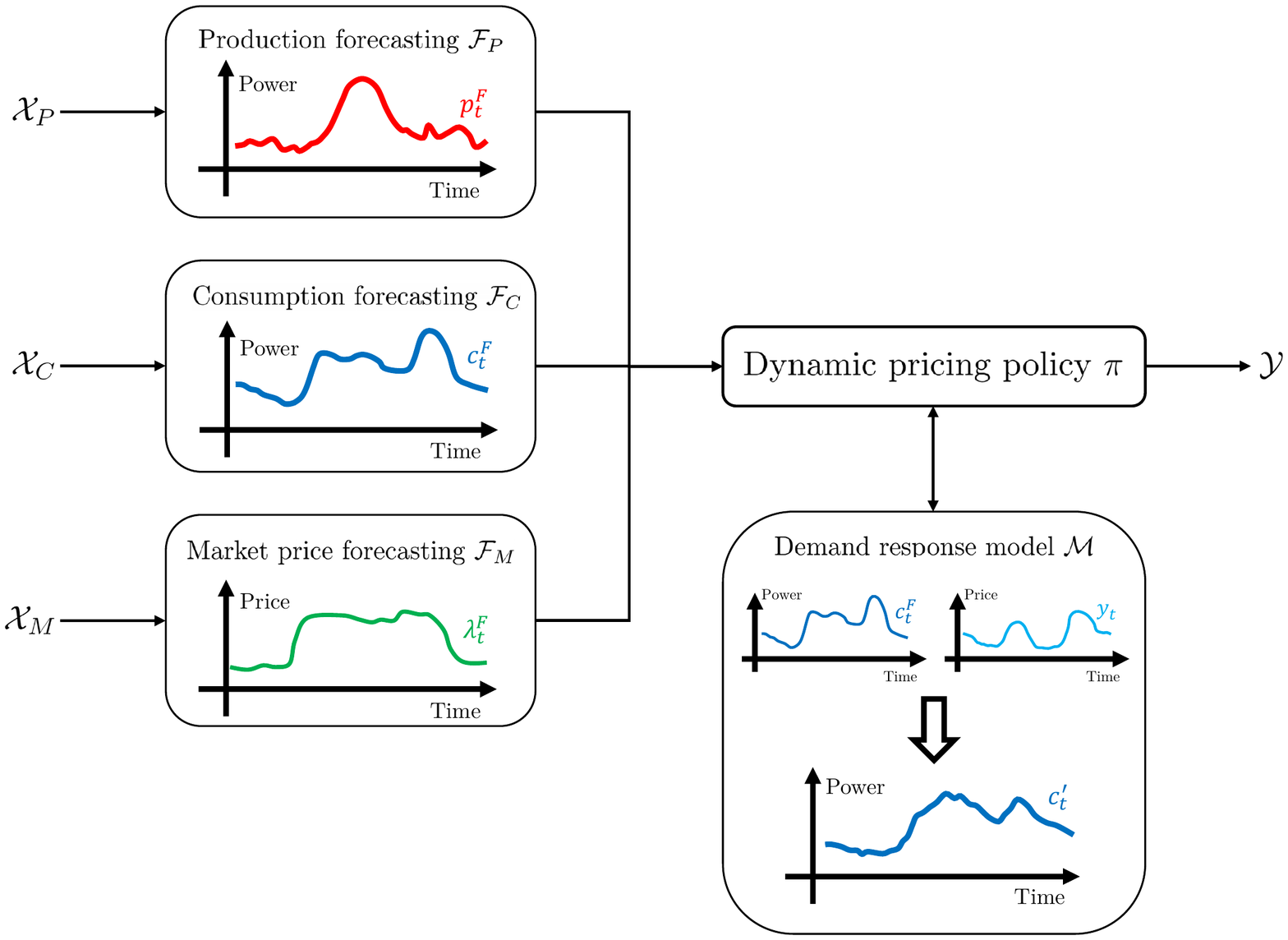}
    \caption{Illustration of the complete decision-making process related to dynamic pricing from the perspective of the supply side, with the connections between the different algorithmic components highlighted.}
    \label{FigureDecisionMakingProcess}
\end{figure*}

In parallel, for the sake of clarity, Figure \ref{FigureDecisionMakingProcess} highlights the interconnections between the different algorithmic components in the scope of a dynamic pricing policy from the perspective of the supply side. Moreover, Algorithm \ref{AlgorithmDecisionMakingProcess} provides a thorough description of the complete decision-making process for the dynamic pricing problem at hand. The complexity of the variable time lag between decision-making and application is highlighted. Assuming that the decision-making occurs once a day at 12:00 AM just before the closing of the day-ahead market for all hours of the following day, the dynamic price at time step $t$ is decided hours in advance at time step $t - \left[12 + (t\%24) \right]$ with the symbol $\%$ representing the modulo operation.

\begin{algorithm*}
\small
\caption{Dynamic pricing complete decision-making process}
\begin{algorithmic}
\STATE \textit{The decision-making occurs once per day before the closing of the day-ahead market at 12:00 AM for all hours of the next day.}
\STATE \textit{The decision-making for the dynamic price of time step $t$ occurs at time step $t - \left[12 + (t\%24) \right]$.}
\FOR{$\tau = -12$ \TO $T-12$}
    \STATE Check whether the time is 12:00 AM to proceed to the decision-making.
    \IF{$(\tau+12) \% 24 = 0$}
        \FOR{$t = \tau+12$ \TO $\tau+36$}
            \STATE Gather the available information for production forecasting $x_{t}^P = \{W_t^F,\ A_t^F,\ I_t^P \}$.
            \STATE Gather the available information for consumption forecasting $x_{t}^C = \{W_t^F,\ T_t,\ I_t^C \}$.
            \STATE Gather the available information for day-ahead market price forecasting $x_{t}^M = \{x_t^P,\ x_t^C,\ G_t^F,\ M_t,\ I_t^M \}$.
            \STATE Forecast production at time step $t$: $p_{t}^F = \mathcal{F}_P \left( x_{t}^P \right)$.
            \STATE Forecast consumption at time step $t$: $c_{t}^F = \mathcal{F}_C \left( x_{t}^C \right)$.
            \STATE Forecast the day-ahead market price at time step $t$: $\lambda_{t}^F = \mathcal{F}_M \left( x_{t}^M \right)$.
        \ENDFOR
        \FOR{$t = \tau+12$ \TO $\tau+36$}
            \STATE Gather the input information for the dynamic pricing policy $x_{t} = \{P_t^{F},\ C_t^{F},\ \Lambda_t^{F},\ Y_t,\ \mathcal{M} \}$.
            \STATE Make a dynamic pricing decision for time step $t$: $y_t = \pi(x_t)$.
        \ENDFOR
        \STATE Announce the dynamic prices for all hours of the following day $\{ y_t \ | \ t = \tau+12, ..., \tau+35 \}$.
    \ENDIF
\ENDFOR
\end{algorithmic}
\label{AlgorithmDecisionMakingProcess}
\end{algorithm*}

\subsection{Production forecasting}
\label{SectionProductionForecasting}

The first forecasting block to be discussed concerns the production of intermittent renewable energy sources such as wind turbines and photovoltaic panels. Indeed, having access to accurate predictions about the future output of the production portfolio is key to the performance of a dynamic pricing policy from the perspective of the supply side. As previously explained in Section \ref{SectionPolicy}, the forecasts have to be available one day ahead before the closing of the day-ahead electricity market for all hours of the following day. Naturally, the generation of such predictions introduces uncertainty, a complexity that has to be taken into account to design sound dynamic pricing policies.\\

Formally, the forecasting model associated with the output of the production portfolio is denoted $\mathcal{F}_P$. Its input space $\mathcal{X}_P$ comprises every piece of information that may potentially have an impact on the generation of electricity from intermittent renewable energy sources such as wind turbines and photovoltaic panels for a certain time period. Its output space $\mathcal{Y}_P$ is composed of a forecast regarding the power generation from the production portfolio for that same time period. Mathematically, the forecasting model input $x_t^P \in \mathcal{X}_P$ and output $y_t^P \in \mathcal{Y}_P$ at time step $t$ can be expressed as follows:
\begin{equation}
    x_t^P = \{W_t^F,\ A_t^F,\ I_t^P \} \text{,}
\end{equation}
\begin{equation}
    y_t^P = p_t^F \text{,}
\end{equation}
where:
\begin{itemize}
    \item [$\bullet$] $W_t^F$ represents various weather forecasts related to the power production of intermittent renewable energy sources such as wind turbines and photovoltaic panels (wind speed/direction, solar irradiance, etc.) at the time step $t$,
    \item [$\bullet$] $A_t^F$ represents predictions about the available capacity of the production portfolio at time step $t$, which may be impacted by scheduled maintenance, repairs, or other similar constraints,
    \item [$\bullet$] $I_t^P$ represents any additional information that may help to accurately forecast the future power generation of the producer/retailer's production portfolio at time step $t$.
\end{itemize}

In the scientific literature, the current state of the art for forecasting the power production of intermittent renewable energy sources is mainly based on deep learning techniques together with some data cleansing processes and data augmentation approaches. The best architectures are \textit{recurrent neural networks} (RNN), \textit{convolutional neural networks} (CNN) and \textit{transformers} \cite{Sweeney2019, Ahmed2020, Aslam2021, Jahangir2021, Heinemann2021}.

\subsection{Consumption forecasting}
\label{SectionConsumptionForecasting}

The objective of the next important forecasting model deserving a discussion is to accurately predict the future power demand of the consumption portfolio before any demand response phenomenon is induced. Since the main goal of a dynamic pricing policy is to maximise the synchronisation between supply and demand, electricity load forecasts are of equal importance to electricity generation predictions. Similarly to the latter, the portfolio consumption forecasts are assumed to be generated one day ahead just before the closing of the day-ahead market for all 24 hours of the following day. Additionally, the uncertainty associated with these predictions has to be seriously taken into account for the success of the dynamic pricing policy.\\

From a more formal perspective, the forecasting model responsible for predicting the future electricity load of the consumption portfolio is denoted $\mathcal{F}_C$. Its input space $\mathcal{X}_C$ includes all the information that may have an influence on the residential electricity consumption for a certain time period. Its output space $\mathcal{Y}_C$ comprises a forecast of the power used by the consumption portfolio for that same time period. Mathematically, the consumption forecasting model input $x_t^C \in \mathcal{X}_C$ and output $y_t^C \in \mathcal{Y}_C$ at time step $t$ can be expressed as the following:
\begin{equation}
    x_t^C = \{W_t^F,\ T_t,\ I_t^C \} \text{,}
\end{equation}
\begin{equation}
    y_t^C = c_t^F \text{,}
\end{equation}
where:
\begin{itemize}
    \item [$\bullet$] $W_t^F$ represents various weather forecasts related to the residential electricity consumption (temperature, hygrometry, etc.) at the time step $t$,
    \item [$\bullet$] $T_t$ represents diverse characteristics related to the time step $t$ (hour, weekend, holiday, season, etc.),
    \item [$\bullet$] $I_t^C$ represents supplementary information that could potentially have an influence on the residential power consumption at time step $t$.
\end{itemize}

Similarly to renewable energy production forecasting, the state-of-the-art approaches for predicting the residential electricity load in the short term are mostly related to deep learning techniques with preprocessed augmented data: RNN, CNN, and transformers \cite{Kong2019, Somu2020, Jin2021, Gasparin2021, Aslam2021}.

\subsection{Market price forecasting}
\label{SectionMarketPriceForecasting}

The last forecasting block to be discussed concerns the future day-ahead electricity market prices. Contrarily to the forecasting of power production and consumption, these price predictions are not critical to the success of a dynamic pricing policy from the perspective of the supply side. Still, having access to quality forecasts for the future day-ahead market prices remains important in order to satisfy the constraints related to the profitability of the producer/retailer as well as the reduced electricity costs for the consumer. Once again, the predictions are assumed to be made just before the closing of the day-ahead market. Moreover, the uncertainty associated with these forecasts has to be taken into consideration.\\

Formally, the forecasting model related to the future day-ahead electricity market prices is denoted $\mathcal{F}_M$. Its input space $\mathcal{X}_M$ includes every single piece of information which may potentially explain the future electricity price on the day-ahead market for a certain hour. Its output space $\mathcal{Y}_M$ comprises a forecast of the day-ahead market price for that same hour. Mathematically, both forecasting model input $x_t^M \in \mathcal{X}_M$ and output $y_t^M \in \mathcal{Y}_M$ at time step $t$ can be expressed as follows:

\begin{equation}
    x_t^M = \{x_t^P,\ x_t^C,\ G_t^F,\ M_t,\ I_t^M \} \text{,}
\end{equation}
\begin{equation}
    y_t^M = \lambda_t^F \text{,}
\end{equation}
where:
\begin{itemize}
    \item [$\bullet$] $G_t^F$ represents forecasts about the state of the power grid as a whole (available production capacity, transmission lines, etc.) at the time step $t$,
    \item [$\bullet$] $M_t$ represents diverse information in various markets related to energy (power, carbon, oil, gas, coal, etc.) in neighbouring geographical areas at time step $t$,
    \item [$\bullet$] $I_t^M$ represents any extra piece of information that may help to predict the future electricity price on the day-ahead market at time step $t$.
\end{itemize}

Once again, the scientific literature reveals that the state-of-the-art approaches for day-ahead power market price forecasting are mostly based on innovative machine learning techniques \cite{Weron2014, Nowotarski2016, Gollou2017, Ugurlu2018, Jahangir2020}.

\subsection{Demand response modelling}
\label{SectionDemandResponseModelling}

Another essential algorithmic component is the mathematical modelling of the residential demand response to dynamic prices. In order to make relevant dynamic pricing decisions, an estimation of the impact of the electricity price on the consumer's behaviour is necessary. In fact, two important characteristics have to be studied:\\

\noindent \textbf{The residential power consumption elasticity.} This quantity measures the average percentage change of the residential power consumption in response to a percentage change in the electricity price. In other words, the elasticity captures the willingness of the consumer to adapt its behaviour when the price of electricity either increases or decreases. This elasticity is critical to the dynamic pricing approach, since it assesses the receptiveness of the consumers to dynamic prices. In fact, the residential power consumption elasticity can be considered as a quantitative indicator of the potential of the dynamic pricing approach.\\

\noindent \textbf{The electricity load temporal dependence.} Time plays an important role in power consumption. Firstly, the consumer's behaviour is highly dependent on the time of the day. The tendency to adapt this behaviour is also expected to be time-dependent. Therefore, the residential power consumption elasticity has to be a function of the time within a day, among other things. Secondly, a higher electricity price does not simply reduce the demand as with other commodities, but rather shifts part of the consumption earlier and/or later in time. This phenomenon reflects a complex temporal dependence for power consumption, which has to be accurately modelled in order to design a performing dynamic pricing policy.\\

Formally, the mathematical model of the residential demand response is denoted $\mathcal{M}$. Its input space $\mathcal{X}_D$ is composed of the predicted power consumption before any demand response and the dynamic prices to be sent to the consumers for several hours before and after the time period analysed, together with information about that time period. Its output space $\mathcal{Y}_D$ comprises the predicted power consumption after demand response to dynamic prices for that same time period. Mathematically, both demand response model input $x_t^D \in \mathcal{X}_D$ and output $y_t^D \in \mathcal{Y}_D$ at time step $t$ can be expressed as the following:
\begin{equation}
    x_t^D = \{C_t^F,\ Y_t',\ T_t \} \text{,}
\end{equation}
\begin{equation}
    y_t^D = c_t' \text{,}
\end{equation}
where $Y_t' = \{y_{t + \epsilon} \in \mathbb{R} \ | \ \epsilon = -k, ..., k\}$ is the dynamic price signal within a time window centred around time step $t$ and of size $k$ from which the demand response is induced.\\

As far as the scientific literature about the modelling of demand response to dynamic prices is concerned, this interesting topic has not yet received a lot of attention from the research community. Still, there exists a few sound works presenting demand response models and assessing the receptiveness of the consumers to dynamic power prices \cite{Gottwalt2011, Bregere2020, Ganesan2022, He2012, Klaassen2016}, as explained in Section \ref{SectionLiterature}.

\subsection{Uncertainty discussion}
\label{SectionUncertainty}

As previously hinted, a dynamic pricing policy has to make its decisions based on imperfect information. Indeed, multiple forecasts for the electricity price, production and consumption have to be generated 12 up to 35 hours in advance. Naturally, these predictions comes with a level of uncertainty that should not be neglected. Moreover, accurately modelling the residential demand response to dynamic prices is a particularly challenging task. Because of both the random human nature and the difficulty to fully capture the consumers' behaviour within a mathematical model, a notable level of uncertainty should also be considered at this stage. Therefore, multiple sources of uncertainty can be identified in the scope of the dynamic pricing decision-making problem at hand, and a proper management of this uncertainty is necessary.\\

A stochastic reasoning is recommended to make sound dynamic pricing decisions despite this substantial level of uncertainty. Instead of considering each uncertain variable (production, consumption, price, demand response) with a probability of 1, the full probability distribution behind these quantities has to be estimated and exploited. Based on this information, the risk associated with uncertainty may be mitigated. Moreover, safety margins may also contribute to reduce this risk, but potentially at the expense of a lowered performance. In fact, there generally exists a trade-off between performance and risk, in line with the adage: \textit{with great risk comes great reward}.

\section{Performance assessment methodology}
\label{SectionPerformanceAssessment}

This section presents a methodology for quantitatively assessing the performance of a dynamic pricing policy in a comprehensive manner. As explained in Section \ref{SectionObjective}, several disjoint objectives can be clearly identified. For the sake of completeness, this research work presents three quantitative indicators, one for each objective. The relative importance of these indicators is left to the discretion of the reader according to its main intention among the different objectives previously defined.\\

The performance indicators proposed are based on the comparison with the original situation without dynamic pricing. In this case, the consumer is assumed to be fully ignorant about the mismatch problem between supply and demand. No information is provided to the customers of the producer/retailer, which consequently have an uninfluenced consumption behaviour. The price of electricity $\overline{e_t}$ is freely determined by the producer/retailer. It may for instance be a fixed tariff, or a price indexed on the day-ahead market price:
\begin{equation}
    \overline{e_t} = \alpha \ \lambda_t + \beta \ \text{,}
\end{equation}
where $\alpha$ and $\beta$ are parameters to be set by the retailer.\\

Firstly, the impact of a dynamic pricing policy on the synchronisation between power supply and demand can be assessed through the performance indicator $S$ quantifying the relative evolution of the deviation $\Delta_T$. This quantity is mathematically expressed as follows:\\
\begin{equation}
    S = 100\ \frac{\overline{\Delta_T} - \Delta_T}{\overline{\Delta_T}}\ \text{,}
\end{equation}
\begin{equation}
    \overline{\Delta_T} = \sum_{t=0}^{T-1} |p_t - \overline{c_t}|\ \text{,}
\end{equation}
where $\overline{\Delta_T}$ represents the lack of synchronisation between supply and demand without dynamic pricing. Therefore, the quantity $S$ has ideally to be maximised, with a perfect synchronisation between supply and demand leading to a value of $100\%$ reduction in deviation.\\

Secondly, the consequence for the consumer regarding its electricity bill can be evaluated with the quantity $B$ which informs about the relative evolution of this power bill. It can be mathematically computed as the following:
\begin{equation}
    B = 100\ \frac{\overline{B_T} - B_T }{\overline{B_T}}\ \text{,}
\end{equation}
where $\overline{B_T} = \sum_{t=0}^{T-1} \overline{c_t}\ \overline{e_t}$ represents the electricity bill paid by the consumer without dynamic pricing.
Since the performance indicator $B$ represents the percentage reduction in costs, it has to ideally be maximised.\\

Lastly, the enhancement in terms of revenue for the producer/retailer can be efficiently quantified thanks to the performance indicator $R$. This quantity represents the relative evolution of the producer/retailer revenue and can be mathematically expressed as follows:
\begin{equation}
    R = 100\ \frac{R_T - \overline{R_T}}{\overline{R_T}}\ \text{,}
\end{equation}
\begin{equation}
    \overline{R_T} = \sum_{t=0}^{T-1} \left[ \overline{c_t}\ \overline{e_t} - (c_t^F - p_t^F) \ \lambda_t - (\overline{c_t} - p_t) \ i_t \right] \text{,}
\end{equation}
where $\overline{R_T}$ represents the producer/retailer revenue without dynamic pricing. Obviously, the performance indicator $R$ has to ideally be maximised.

\section{Conclusion}
\label{SectionConclusion}

This research paper presents a detailed formalisation of the decision-making problem faced by a producer/retailer willing to adopt a dynamic pricing approach, in order to induce an appropriate residential demand response. Three core challenges are highlighted by this formalisation work. Firstly, the objective criterion maximised by a dynamic pricing policy is not trivially defined, since different goals that are not compatible can be clearly identified. Secondly, several complex algorithmic components are necessary for the development of a performing dynamic pricing policy. One can for instance mention different forecasting blocks, but also a mathematical model of the residential demand response to dynamic prices. Thirdly, the dynamic pricing decisions have to be made based on imperfect information, because this particular decision-making problem is highly conditioned by the actual uncertainty for the future.\\

Several avenues are proposed for future work. In fact, the natural extension of the present research is to design innovative dynamic pricing policies from the perspective of the supply side based on the formalisation performed. While the present research paper exclusively focuses on the philosophy and conceptual analysis of the approach, there remain practical concerns that need to be properly addressed in order to achieve performing decision-making policies. To achieve that, a deeper analysis of the scientific literature about each algorithmic component discussed in Section \ref{SectionDiscussion} is firstly welcomed, in order to identify and reproduce the state-of-the-art techniques within the context of interest. Then, different approaches have to be investigated for the design of the dynamic pricing policy itself. One can for instance mention, among others, the stochastic optimisation and deep reinforcement learning techniques. Finally, the dynamic pricing policies developed have to be rigorously evaluated, analysed, and compared by taking advantage of real-life experiments.

\section*{Acknowledgements}

Thibaut Théate is a Research Fellow of the F.R.S.-FNRS, of which he acknowledges the financial support.

\bibliography{manuscript}

\end{document}